\documentstyle[multicol,aps]{revtex}
\begin{document}
\draft
\title{Aharonov-Bohm effect in the chiral Luttinger liquid}
\author{Michael R. Geller$^1$ and Daniel Loss$^2$}
\address{$^1$Department of Physics,
Simon Fraser University, Burnaby B.C. V5A 1S6, Canada\\
$^2$Department of Physics, University of Basel, Klingelbergstrasse 82,
4056 Basel, Switzerland}
\date{\today}
\maketitle

\begin{abstract}
Edge states of the quantum Hall fluid provide an almost unparalled
opportunity to study mesoscopic effects in a highly correlated electron
system. In this paper we develop a bosonization formalism for the
finite-size edge state, as described by chiral Luttinger liquid theory,
and use it to study the Aharonov-Bohm effect. The problem we address
may be realized experimentally by measuring the tunneling current
between two edge states through a third edge state
formed around an antidot in the fractional quantum Hall effect regime.
The finite size $L$
of the antidot edge state introduces a temperature scale
$T_0 \equiv \hbar v / \pi k_B L$, where $v$ is the edge-state Fermi velocity.
A renormalization group analysis reveals the existence of a two-parameter
universal scaling function ${\tilde G}(X,Y)$ that describes the
Aharonov-Bohm resonances.
We also show that the strong renormalization of the tunneling
amplitudes that couple the
antidot to the incident edge states, together with the nature of the
Aharonov-Bohm interference process in a chiral system, prevent the
occurrence of perfect resonances as the magnetic field is varied, even
at zero temperature.
In an experimentally realizable strong-antidot-coupling regime,
where the source-to-drain transmission is weak, and at bulk
filling factor $g = 1/q$ with $q$ an odd integer, we predict the
low-temperature $(T \ll T_0)$ Aharonov-Bohm amplitude to
vanish with temperature as $T^{2q-2}$, in striking contrast to
a Fermi liquid $(q=1)$. Near $T_0$, there is a
pronounced maximum in the amplitude, also in contrast to a Fermi
liquid. At high temperatures $(T \gg T_0)$, however, we predict a
crossover to a $T^{2q-1} e^{-q T/T_0}$ temperature dependence,
which is qualitatively similar to chiral Fermi liquid behavior.
Careful measurements in the strong-antidot-coupling regime
above $T_0$ should be able to distinguish between a Fermi liquid
and our predicted
nearly Fermi-liquid scaling. In addition, we predict
an interesting high-temperature nonlinear response
regime, where the voltage satisfies $V > T > T_0$, which
may also be used to distinguish between chiral Fermi liquid and chiral
Luttinger liquid behavior. Finally, we predict new mesoscopic
edge-current oscillations, which are similar to the persistent current
oscillations in a mesoscopic ring, except that they are not reduced
in amplitude by weak disorder. In the fractional quantum Hall regime,
these ``chiral persistent currents'' have a universal
non-Fermi-liquid temperature dependence and may be another ideal
system to observe a chiral Luttinger liquid.
\end{abstract}

\begin{multicols}{2}

\section{introduction}

It is well known that the integer quantum Hall
effect\cite{Prange and Girvin} and many related transport
phenomena that occur in nanometer-scale semiconductor
devices\cite{Beenakker review} can be understood in
terms of Fermi liquid theories
of magnetic edge-state transport\cite{Halperin}, such as the
B\"uttiker-Landauer formalism\cite{BL}. However, as was
shown by Laughlin\cite{Laughlin}, the fractional quantum Hall
effect (FQHE) occurs because strong
electron-electron interactions result in the formation of
highly correlated incompressible states at certain Landau level filling
factors \cite{Prange and Girvin}.
In 1990, Wen\cite{Wen proposal} used the Chern-Simons 
theory\cite{Chern-Simons theory} of the bulk FQHE to show that the
edge states in the fractional regime should be chiral Luttinger
liquids (CLL). As in the nonchiral Luttinger liquid\cite{LL,Haldane},
electron-electron interactions in the CLL play an essential role and
lead to physical properties that can be dramatically different
than the integral quantum Hall effect edge state. Wen's important proposal has
stimulated a considerable theoretical effort to understand the
properties of this non-Fermi-liquid state of matter\cite{Wen reviews,KF
LL,Chamon and Wen,Moon etal,Pokrovsky and Pryadko,KFP,noise,Fendley etal,Oreg
and Finkelstein,KF CLL,Renn and Arovas,Balatsky and Matveenko,Palacios
and MacDonald,Moon and Girvin,Geller etal,Chamon etal}.

The first experimental observation of a CLL was made by
Milliken, Umbach, and Webb \cite{Milliken etal}. These authors
measured the tunneling current between two filling factor $1/3$ edge states in
a quantum-point-contact geometry.
As the gate voltage was varied,
resonance peaks in the conductance, caused by conditions
of destructive interference that prevent tunneling between the edges,
were observed to have the
correct CLL temperature dependence as predicted by Moon and
coworkers \cite{Moon etal}
and by Fendley, Ludwig, and Saleur \cite{Fendley etal}. In addition,
Chang, Pfeiffer, and
West \cite{Chang etal},
working with a cleaved-edge sample and measuring the tunneling between
a single filling factor $1/3$ edge state and doped GaAs, have very
recently reported experimental
evidence that is also in favor of CLL theory.
Thus, it would appear that the CLL is indeed realized in FQHE edge states.

However, two recent experiments \cite{Franklin etal,Maasilta and Goldman}
on the tunneling between FQHE edge states through an additional edge state
circling a quantum antidot have reported Fermi liquid behavior.
In the quantum-point-contact geometry the tunneling occurs through an
unknown and uncontrollable impurity distribution, resulting in a dense
series of resonance peaks \cite{Milliken etal}.
In contrast, the tunneling in the anitdot
system occurs through a larger object, essentially a mesoscopic
ring, and the resonances are controlled by the Aharonov-Bohm (AB) effect by
varying the magnetic field within a given FQHE plateau. For this reason
the antidot geometry is believed to be superior for observing
resonant tunneling in a CLL.
Franklin {\it et al.}\cite{Franklin etal} measured the AB
conductance oscillations in a device with an antidot $0.94 \ \mu {\rm m}$
in circumference at temperatures down to $30 \ {\rm mK}$. The filling
factor in the immediate vicinity of the antidot was measured to be
$1/3$, whereas in the bulk it was $2/3$. In both the
integer and fractional regimes the period of the AB oscillations
was observed to correspond to one flux quantum through the area
bounded by the antidot edge state.
The temperature dependence was analyzed by determining the temperature
dependence of the appropriate Fourier coefficient of the transformed
resistance data,
and a good fit to Fermi liquid scaling was obtained.
More recently, Maasilta and Goldman\cite{Maasilta and Goldman}, following
earlier related work by Goldman and Su\cite{Goldman and Su},
studied resonant tunneling
as a function of back-gate voltage down to $12 \ {\rm mK}$ in an
antidot $1.9 \ \mu {\rm m}$ in circumference. Within the experimental
uncertainty, the resonance lineshape at a given temperature was
consistent with both chiral Fermi liquid and CLL theory. The temperature
dependence of the width and amplitude of the oscillations, however, was
found to be that of a chiral Fermi liquid.

The agreement of these experiments with Fermi liquid theory does not in
itself rule out CLL theory because no detailed CLL theory for the antidot
geometry has been available. This is one motivation of the present work.
The most important conclusion of our study is that the transport
properties of the quantum-point-contact system and the antidot system
differ in two important ways. The first is that mesoscopic effects are
very important in the latter. When the thermal length
$L_{\rm T} \equiv \hbar v/ k_{\rm B} T$ becomes smaller than the
circumference $L$ of the antidot edge state, the AB oscillations become
washed out, and, at the same time, acquire a temperature dependence
that is similar to a chiral Fermi liquid \cite{Geller etal}.
Here $v$ is the edge-state Fermi velocity.
An experiment
performed at a temperature
significantly above the point of crossover,
\begin{equation}
T_0 \equiv {\hbar v \over \pi k_{\rm B} L},
\label{T0}
\end{equation}
is therefore {\it expected} to observe nearly Fermi liquid behavior
for many mesoscopic quantities.
(The precise definition of $T_0$ has been chosen to simplify the
expressions in Section V.)
The second difference is that in contrast with the quantum-point-contact
geometry, where it is reasonable to assume that there exist conditions of
destructive interference that lead to perfect resonances, the resonances
in the antidot geometry, which are controlled by the AB effect, are never
perfect, even at zero temperature. As we shall explain, this means that
there is another temperature scale $T_1$, set by the 
bare tunneling rate between
the antidot and incident edge states, below which the system is always in the
strongly coupled regime.

The other motivation for our work is that the CLL provides an
almost unparalled opportunity to study mesoscopic physics in a highly
correlated electron system that is both experimentally accessible
and theoretically tractable. Thus, our comparison of the AB
effect in the Fermi and Luttinger liquids is also a comparison of
mesoscopic effects in a noninteracting and interacting system, and
we shall show that at temperatures below $T_0$ interactions have a
dramatic effect on many quantities. Mesoscopic effects in nonchiral
Luttinger liquids have been studied considerably
\cite{Haldane,Loss,mesoscopic LL},
and a recent interesting paper by Chamon and coworkers \cite{Chamon etal}
has analyzed mesoscopic effects in a CLL by considering a double point-contact
arrangement that allows one to measure the fractional charge and
fractional statistics of FQHE quasiparticles. Although the main focus
of their work is different, and the geometry they consider is close to
that of a quantum dot instead of an antidot, many of the results of
Chamon {\it et al.} \cite{Chamon etal} are similar to ours.

The model of the antidot system we shall adopt for our study is the
simplest one possible: We assume two macroscopic filling factor $g = 1/q$
(with $q$ odd) edge states at the edges of the Hall bar symmetrically
coupled to a single mesoscopic edge state (with the same $g$) circling
the antidot. The edges of the Hall fluid are assumed to be sharply
confined, and the interaction short-ranged, so that the low lying
excitations consist of a single branch of edge-magnetoplasmons with
linear dispersion $\omega = v k$. It is not known at present whether
or not the deviations of the experimentally studied systems from this
idealized model are important.

The original effect of Aharonov and Bohm \cite{Aharonov and Bohm}
was proposed as an experiment with electron beams in a vacuum,
but it was realized immediately that electrons moving in a
multiply connected conductor should  also be influenced by a
magnetic flux threaded through it. Because it is a sensitive
probe of phase coherence, the AB effect has been an
important experimental tool to study disordered
metals \cite{Aronov and Sharvin}.
Before proceeding with our study,
it is worthwhile to inquire whether there are
any essential differences between the AB effect in nonchiral and chiral
systems. The answer is {\it yes}: In nonchiral systems, the AB flux can
lead to both constructive and destructive
interference, whereas only constructive interference
is possible in the chiral system. This difference can be understood
in the following way.

Let $\theta_\pm \equiv \oint_\epsilon d{\bf l} \cdot
({\bf p} + {e \over c} {\bf A})$
be the phase accumulated by an electron with energy
$\epsilon$ after one complete clockwise $(+)$ or counterclockwise $(-)$ orbit
around the ring shown in Fig.~\ref{ring}; it is given by
$\theta_\pm = 2 \pi ( {\textstyle{\epsilon \over \Delta \epsilon}} \pm
\varphi)$
and includes both the dynamical and AB phases.  Here $e$ is the magnitude
of the
electron charge, $\varphi \equiv \Phi / \Phi_0$
is the dimensionless AB flux threading the ring, $\Phi_0 \equiv hc/e$ is the
flux quantum, and
$\Delta \epsilon \equiv 2 \pi v/L$ is the energy level spacing for
noninteracting
electrons having linear dispersion with velocity $v$ in a ring of
circumference $L$.
The probability amplitude ${\cal A}_\pm$ to propagate on the ring from
point 1 to point 2 at
energy $\epsilon$, allowing only clockwise $(+)$ or counterclockwise $(-)$
motion, is given by
\begin{equation}
{\cal A}_\pm = e^{i \theta /2} + e^{i 3\theta /2} + e^{i 5\theta /2} +
\cdots =
{ i/2 \over
\sin \pi({\textstyle{\epsilon \over \Delta \epsilon}} \pm \varphi) }.
\end{equation}
The first term in the series, $e^{i \theta /2}$, is the amplitude to
propagate directly from point 1 to point 2, and the remaining terms account
for any number of windings with a given chirality. The total 
\lq\lq transmission\rq\rq
probability in the chiral case is therefore
\begin{equation}
|{\cal A}_\pm|^2 = { 1/2 \over
1 - \cos 2 \pi({\textstyle{\epsilon \over \Delta \epsilon}} \pm \varphi) },
\end{equation}
which possesses transmission resonances when
$ {\textstyle{\epsilon \over \Delta \epsilon}} \pm \varphi$ is integral,
but never exhibits completely destructive interference.
In contrast, the total transmission probability in the nonchiral case,
\begin{equation}
|{\cal A}_{+} + {\cal A}_{-}|^2 =
{ \sin^2 (\pi {\textstyle{\epsilon \over \Delta \epsilon}}) \ \cos^2 (\pi
\varphi)
\over \sin^2 \pi ({\textstyle{\epsilon \over \Delta \epsilon}} + \varphi)
\sin^2 \pi ({\textstyle{\epsilon \over \Delta \epsilon}} - \varphi)},
\end{equation}
has both poles and zeros.
A more precise analysis of this distinction, based on a comparison of the
chiral and nonchiral propagators, is provided in Appendix A.

The organization of this paper is as follows. In Section II we study
the AB effect in the chiral Fermi liquid for arbitrary
antidot-coupling strength with the B\"uttiker-Landauer
formula. In Section III we
discuss the general theory of the finite-size CLL, including canonical
quantization and bosonization. We also show there how fractionally charged
excitations arise naturally in the CLL. Section IV is devoted to a
renormalization group analysis of the weak-antidot-coupling
regime, where we obtain the same flow equations
as previously derived for the quantum-point-contact system.
The strong-antidot-coupling regime of the AB effect
in a CLL is
studied in detail in Section V. In Section VI we study the response
of an edge state to an AB flux and find
mesoscopic edge-current oscillations that are similar
to persistent current oscillations in a mesoscopic ring, except that
they are not degraded by weak disorder. Section VII contains a brief
discussion of our results and their relevance to the existing
antidot experiments.

\section{aharonov-bohm effect in the chiral fermi liquid}

We begin by studying the chiral Fermi liquid case with the B\"uttiker-Landauer
formula, which is valid for noninteracting electrons.
In addition to serving as a check of our more general
expressions derived below, the B\"uttiker-Landauer formula is valid
for arbitrary antidot-coupling amplitude and the resulting conductance
exhibits
resonances that are not accessible from a perturbation
expansion valid for small tunneling. The B\"uttiker-Landauer analysis also
shows
that the nature of the AB resonances in the chiral Fermi liquid
and CLL are entirely different:
In the weak-antidot-coupling regime
of Fig.~\ref{geometry}a,
with tunneling amplitudes $|\Gamma_{\pm}| \ll 1$,
the two-terminal conductance in
the Fermi liquid possesses reflection resonances (sharp dips in the
conductance) when the electron going around the antidot interferes
{\it constructively} with itself, whereas, as we shall show below, the
chiral Luttinger liquid instead exhibits resonant transmission
(sharp peaks in the conductance)
at conditions of maximum {\it destructive} interference.

Transport through an antidot in the integer quantum Hall effect
regime has been studied both
theoretically\cite{Kirczenow antidot,Mace etal,Tan
and Inkson} and experimentally\cite{IQHE antidot experiments}.
In the two-terminal B\"uttiker-Landauer approach,
the constriction containing the antidot is regarded as a
single phase-coherent scatterer connecting perfect reservoirs serving both as
current sources and voltage probes. The current flowing from reservoir
$1$ to reservoir $2$ is
\begin{equation}
I = - {e \over h} \int d\epsilon \ T(\epsilon)
\ \! \big[ n_{\rm F}(\epsilon - \mu_1) -
n_{\rm F}(\epsilon - \mu_2) \big] ,
\label{LB formula}
\end{equation}
where $T(\epsilon)$ is the total probability for transmission from reservoir
$1$ to $2$, $n_{\rm F}(\epsilon) \equiv (e^{\beta \epsilon} +1)^{-1}$
is the Fermi distribution function, and $\mu_i$ is the electrochemical
potential of reservoir $i$. The voltage $V \equiv (\mu_2 - \mu_1)/e$ is
defined so that a positive $V$ produces a positive $I$.
The two-terminal linear conductance is
\begin{equation}
G = - {e^2 \over h} \int d\epsilon \ {\partial n_{\rm F} \over
\partial \epsilon} \ T(\mu + \epsilon) ,
\label{LB G}
\end{equation}
where $\mu \equiv {\textstyle{\mu_1 +\mu_2 \over 2}}$ is the mean
electrochemical
potential, which reduces to
$G=T(\mu) \ \! {e^2 \over h}$
at zero temperature.

The transmission probability $T(\epsilon)$ has been evaluated for a
variety of edge state configurations in Ref. \cite{Kirczenow antidot}.
For our purposes it is sufficient to consider only the case where
$\Gamma_{\pm} = i \Gamma$ with $\Gamma$ real and energy independent
\cite{tunneling amplitude footnote}.
In this case the amplitude
to tunnel on or off the antidot is $ i \Gamma$, whereas, by unitarity,
the amplitude to proceed without tunneling is
$\sqrt{1-\Gamma^2}$.
Then for the system shown in Fig.~\ref{geometry}a it is simple
to show that
\begin{equation}
T(\epsilon) = 1 - { \Gamma^4 \over
2(1-\Gamma^2)[1- \cos\theta(\epsilon)] + \Gamma^4},
\label{transmission}
\end{equation}
where $\theta = 2 \pi ({\textstyle{\epsilon \over \Delta \epsilon}} +
\varphi)$
is the phase shift of the electron wave function after a complete clockwise
orbit around the antidot at energy $\epsilon$, as defined in Section I, with
$\varphi$ now the dimensionless magnetic flux through the area defined by the
antidot edge state. (Note that the arrows in Fig.~\ref{geometry} denote
the flow of {\it currents}; the electrons are therefore circling the
antidot in
the clockwise direction.)

In the weak-antidot-coupling regime, where $\Gamma \ll 1$,
(\ref{transmission}) shows that sharp zero-temperature reflection
resonances in the two-terminal conductance occur when
$\theta = 2 \pi n $ with $n$ an integer; that is, at
conditions of constructive interference.
In the strong-antidot-coupling regime, where the tunneling
amplitude $\Gamma$ is close to unity, the source-drain transmission
probability is small. In this regime it is convenient to
define a new small parameter
${\bar \Gamma} \equiv \sqrt{1 - \Gamma^2}.$
Then in the strong-coupling
limit $(\bar \Gamma \ll 1)$ we have
\begin{equation}
T(\epsilon) = 2 {\bar \Gamma}^2 [1-\cos\theta(\epsilon)]
+ {\cal O}({\bar \Gamma}^4).
\label{strong-coupling transmission}
\end{equation}
Note that reflection resonances still occur when $\theta = 2 \pi n$,
although they are less
sharp than in the weak-antidot-coupling limit.
This limit corresponds to the case for which the
CLL theory
described in Section V has been developed.
The zero-temperature AB resonances in the weak and strong coupling regimes
are shown in Fig.~\ref{AB resonances}.

An analytic expression for the linear and nonlinear response in the
strong-antidot-coupling regime can be obtained from (\ref{LB formula})
and (\ref{strong-coupling transmission}). The required integral,
\begin{eqnarray}
&\int_{-\infty}^\infty&  d\epsilon
\  { \cos 2\pi \big({\textstyle{\epsilon \over \Delta \epsilon}}
+{\textstyle{\mu_1 \over \Delta \epsilon}} + \varphi \big)
- \cos 2\pi \big({\textstyle{\epsilon \over \Delta \epsilon}}
+{\textstyle{\mu_2 \over \Delta \epsilon}} + \varphi \big)
\over e^{\beta \epsilon} + 1} \nonumber \\
&=& -2 \pi T \ { \sin \big({\textstyle{eV \over 2 \pi T_0}}\big)
\cos 2\pi \big({\textstyle{\mu \over \Delta \epsilon}} + \varphi \big)
\over \sinh \big({\textstyle{T \over T_0}} \big)},
\end{eqnarray}
follows from a contour integration and residue summation.
Here we have used the definitions of $\mu$ and $V$ to write
$\mu_1 = \mu - eV/2$ and $\mu_2 = \mu + eV/2$.
In this regime we therefore find for the chiral Fermi liquid
\begin{equation}
I^{\rm FL} = I_0^{\rm FL} + I^{\rm FL}_{\rm AB}
\ \! \cos 2 \pi \big({\textstyle{\mu \over \Delta \epsilon}} + \varphi \big),
\label{I FL}
\end{equation}
where
\begin{equation}
I^{\rm FL}_0 = { e^2 {\bar \Gamma}^2 \over \pi \hbar} V
\label{FL I0}
\end{equation}
is the flux-independent contribution, and
\begin{equation}
I^{\rm FL}_{\rm AB} = - {2 e {\bar \Gamma}^2 \over \hbar}
{T \over \sinh(T/T_0)}
\sin \bigg({ eV \over 2 \pi T_0 } \bigg)
\label{FL IAB}
\end{equation}
is the AB contribution.
If the voltage $V$ is applied {\it symmetrically} about an antidot energy
level then $\mu$ is an integral multiple of $\Delta \epsilon$ and the voltage 
dependence in the second term of (\ref{I FL}) becomes simply
$\sin(eV/2 \pi T_0) = \sin(eV \pi / \Delta \epsilon)$. The voltage
dependence is then sinusoidal with a period equal to {\it twice} the
antidot level spacing $\Delta \epsilon$, because as the voltage is
varied the two chemical potentials $\mu_1$ and $\mu_2$ move in opposite
directions at half the rate at which $V$ changes.
The linear conductance
$G \equiv (dI/dV)_{V=0}$ in this regime is given by
\begin{equation}
G^{\rm FL} = G_0^{\rm FL} + G^{\rm FL}_{\rm AB}
\ \! \cos 2 \pi \big({\textstyle{\mu \over \Delta \epsilon}} + \varphi \big),
\label{G FL}
\end{equation}
where
\begin{equation}
G^{\rm FL}_0 = { e^2 {\bar \Gamma}^2 \over \pi \hbar} V
= 2 {\bar \Gamma}^2 {e^2 \over h}
\label{FL G0}
\end{equation}
and
\begin{equation}
G^{\rm FL}_{\rm AB} = - {e^2 {\bar \Gamma}^2 \over \pi \hbar}
{T/T_0 \over \sinh(T/T_0)}
= - 2 {\bar \Gamma}^2 {T/T_0 \over \sinh(T/T_0)} {e^2 \over h}.
\label{FL GAB}
\end{equation}
The factor of $2$ in the background term (\ref{FL G0}) comes from
the two parallel
tunneling paths in Fig.~\ref{geometry}b, each having transmission probability
${\bar \Gamma}^2$.
The expressions (\ref{I FL}) and (\ref{G FL}) show that the line shape of
the AB
oscillations as a function of flux or $\mu$ is strictly sinusoidal, with a
temperature-independent linewidth. Only the amplitude of the oscillations,
given
by $I_{\rm AB}^{\rm FL}$ and $G_{\rm AB}^{\rm FL}$, has a
temperature dependence. Note that
(\ref{G FL}) also shows that the relevant Fourier component of the
conductance oscillations has the same temperature dependence as
$G_{\rm AB}^{\rm FL}$.

The calculation presented above is valid for a noninteracting
electron system only, and therefore does not apply to FQHE
edge states. Nonetheless, it is possible to extend these Fermi
liquid results to the FQHE regime by assuming that the
B\"uttiker-Landauer formula (\ref{LB formula}) and the transmission probability
(\ref{transmission}) can be applied to {\it noninteracting}
composite fermions \cite{Kirczenow CF}. With this assumption
the transport properties in the FQHE regime become qualitatively similar
to that in the integer regime and to the existing antidot
experiments \cite{Franklin etal,Maasilta and Goldman}.
The approach we take in this paper, however, is the microscopic
one based on CLL theory.

\section{finite-size chiral Luttinger liquid with topological excitations}

To study mesoscopic effects associated with edge states in the FQHE,
we shall perform a quantization of CLL theory
for a finite-size system and include the possibility of
topological excitations of
the chiral scalar field and coupling to an AB flux.
Finite-size effects in nonchiral Luttinger
liquids have been discussed previously by Haldane \cite{Haldane} and by
Loss \cite{Loss}. To proceed
in the chiral case we bosonize the electron field operators
$\psi_{\pm}(x)$ according to the convention
\begin{equation}
\rho_\pm(x) = \pm { \partial_x \phi_{\pm} \over 2 \pi} ,
\label{bosonization convention}
\end{equation}
where
\begin{equation}
\rho_{\pm}(x) \equiv \lim_{a \rightarrow 0} : \psi_\pm^\dagger(x+a)
\psi_\pm(x):
\end{equation}
is the normal-ordered charge density and $\phi_\pm(x)$ is a chiral scalar
field for right (+) or left (--) movers. The dynamics
is governed by Wen's Euclidian action \cite{Wen reviews}
\begin{equation}
S_\pm ={1\over 4\pi g} \int_0^L \! \! dx \int_0^\beta \! \! d\tau
\ \partial_x \phi_\pm
\big(\pm i\partial_\tau \phi_\pm + v \partial_x \phi_\pm \big),
\label{euclidian action}
\end{equation}
where $g=1/q$ (with $q$ odd) is the bulk filling factor, $v$ is the
edge-state Fermi velocity (edge-magnetoplasmon velocity), and $L$
is the size of the edge state.
When $q=1$, the action (\ref{euclidian action}) describes noninteracting
chiral electrons.
The Lagrangian and real-time equations of motion are
\begin{equation}
{\cal L}_\pm = {1 \over 4 \pi g} \ \!  \partial_x \phi_\pm
\big( \mp \partial_t \phi_\pm - v \partial_x \phi_\pm \big),
\label{lagrangian}
\end{equation}
and
\begin{equation}
\big(\partial_x \partial_t \pm v \partial_x^2 \big) \phi_\pm(x,t) = 0.
\label{field equations}
\end{equation}
The field theory described by (\ref{euclidian action})
can be canonically quantized by imposing the equal-time commutation
relation
\begin{equation}
[ \phi_{\pm}(x) , \phi_{\pm}(x')] = \pm i \pi g \ \! {\rm sgn}(x-x').
\label{full commutation relation}
\end{equation}
Furthermore, the left and right sectors commute,
\begin{equation}
[\phi_{-}(x), \phi_{+}(x')] = 0.
\end{equation}
The momentum density canonically conjugate to $\phi_\pm(x)$
is therefore identified as
$\mp \partial_x \phi_\pm / 2 \pi g$.
We then decompose $\phi_{\pm}(x)$ into a nonzero-mode contribution
$\phi_\pm^{\rm p}(x)$ satisfying periodic boundary conditions that
describes the neutral excitations, and a zero-mode part
$\phi_{\pm}^0(x)$ that contributes to the charged excitations,
\begin{equation}
\phi_{\pm}(x) = \phi_\pm^{\rm p}(x) + \phi_{\pm}^0(x).
\label{decomposition}
\end{equation}
The nonzero-mode part may be expanded in a basis of Bose
annihilation and creation operators as
\begin{equation}
\phi_{\pm}^{\rm p}(x) = \sum_{k \neq 0} \theta(\pm k)
\sqrt{\textstyle{2 \pi g \over |k| L}}
\big( a_k e^{ikx} + a_k^\dagger e^{-ikx} \big) e^{-|k|a/2} ,
\label{nonzero-mode expansion}
\end{equation}
where $[a_k, a_{k'}^\dagger] = \delta_{k k'}$,
with the coefficients in (\ref{nonzero-mode expansion})
determined by the requirement that
$\phi_{\pm}^{\rm p}(x)$ itself satisfies
(\ref{full commutation relation}) in the
$L \rightarrow \infty$ limit.
In a finite-size system, however, it can be shown that
\begin{equation}
[\phi_\pm^{\rm p}(x),\phi_\pm^{\rm p}(x')]
= \pm i \pi g \ \! {\rm sgn}(x-x') \mp {2 \pi i g (x-x') \over L},
\label{nonzero-mode commutation relation}
\end{equation}
so we must require the zero-mode contribution to satisfy
\begin{equation}
[\phi_{\pm}^0(x), \phi_{\pm}^0(x')] = \pm{2 \pi i g (x-x') \over L}
\label{zero-mode commutation relation}
\end{equation}
for the total field to satisfy (\ref{full commutation relation}).
An expansion analogous to (\ref{nonzero-mode expansion}) for
the zero-modes can be constructed from the condition
(\ref{zero-mode commutation relation}) and, in addition, the requirement
\begin{equation}
\phi_{\pm}^0(x+L) - \phi_\pm^0(x) = \pm 2 \pi N_\pm ,
\label{zero-mode boundary condition}
\end{equation}
which follows from (\ref{bosonization convention}), where
\begin{equation}
N_\pm \equiv \int_0^L dx \ \rho_{\pm}(x)
\end{equation}
is the charge of an excited
state relative to the ground state.
Conditions (\ref{zero-mode commutation relation}) and
(\ref{zero-mode boundary condition}) together determine $\phi_\pm^0(x)$,
up to an additive c-number constant, as
\begin{equation}
\phi_\pm^0(x) = \pm {2 \pi \over L} N_\pm x - g \ \! \chi_\pm ,
\label{zero-mode expansion}
\end{equation}
where $\chi_\pm$ is an Hermitian phase operator canonically conjugate to
$N_\pm$ satisfying
\begin{equation}
[\chi_\pm, N_\pm] = i.
\end{equation}
Equations (\ref{nonzero-mode expansion}) and (\ref{zero-mode expansion})
may now be used to write the normal-ordered CLL Hamiltonian as
\begin{eqnarray}
H_\pm &=& {v \over 4 \pi g} \int_0^L dx \big(\partial_x \phi_\pm \big)^2 \\
&=& {\pi v \over g L} N_\pm^2 + \sum_{k} \theta(\pm k) v |k|
a_k^\dagger a_k .
\label{hamiltonian}
\end{eqnarray}
The normal-ordered charge $N_\pm$ is a constant of the motion, whereas
$\partial_t \chi_\pm = 2 \pi v N_\pm / gL$.
The normal-ordered density operator (\ref{bosonization convention}) for an
isolated edge state satisfies the chiral equations of motion
\begin{equation}
\big( \partial_t \pm v \partial_x \big) \rho_{\pm} = 0.
\label{chiral equations of motion}
\end{equation}
Also note that the
compressibility $\kappa_\pm \equiv \partial \rho_\pm/\partial \mu$
of the uniform CLL is $\kappa_\pm = g/2 \pi v$, half the
spinless nonchiral Luttinger liquid value.
In a finite-size system, the level spacing for neutral and charged
excitations is of the order of $v/L$, and both types of edge excitations
become gapless in
the $L \rightarrow \infty$ limit as expected.

We turn now to a discussion of the bosonization of the right $(+)$
and left $(-)$ moving components of the electron field operators.
Eqn.~(\ref{bosonization convention}) shows that to create an electron,
we need to create a $\pm 2 \pi$ step in the chiral scalar field.
The electron field operators can be bosonized as
\begin{equation}
\psi_\pm(x) = {1\over \sqrt{2 \pi a}} \,
e^{i [\phi_\pm(x)  \pm {\pi x \over L }]/g},
\label{bosonization}
\end{equation}
where $a$ is the same microscopic cutoff length that appears in
(\ref{nonzero-mode expansion}).  To see that (\ref{bosonization}) is
valid, note that
\begin{equation}
[\rho_\pm(x) , \psi_\pm^\dagger(x')]
= \delta(x-x') \psi_\pm^\dagger(x'),
\end{equation}
so $\psi_\pm^\dagger(x)$ creates an electron at position $x$.
The necessary time-dependent generalization of (\ref{bosonization}) is
\begin{equation}
\psi_{\pm}(x,t) = {1 \over \sqrt{2 \pi a}} e^{i \phi_{\pm}(x,t)/g}
e^{\pm i \pi (x\mp vt) / gL},
\end{equation}
where $\phi_{\pm}(x,t)$ is in the Heisenberg representation.

It is important to realize that the additional c-number
phase factor $e^{\pm i \pi x/gL}$, which has the effect of
disentangeling the charge and phase operators in the zero-mode,
is necessary for bosonizaton
in a finite-size system. An example of this necessity is
provided by a calculation of the equal-time correlation
function
\begin{equation}
C_\pm(x) \equiv \big\langle {\psi_\pm}(x)
{\psi_\pm^\dagger}(0) \big\rangle
\label{correlator definition}
\end{equation}
for the finite-size chiral Fermi liquid $(g=1)$ at
zero temperature, which can be calculated via bosonization
and also directly. Using (\ref{bosonization}) we find
\begin{eqnarray}
C_\pm(x) &=& {e^{\pm i \pi x/L} \over 2 \pi a}
\big\langle e^{i({\phi_\pm^0}(x) - {\phi_\pm^0}(0) )} \big\rangle \nonumber \\
& \times & e^{{1 \over 2}[\phi_\pm^0(x),\phi_\pm^0(0)]}
e^{\langle {\phi_\pm^{\rm p}}(x) {\phi_\pm^{\rm p}}(0)
- ({\phi_\pm^{\rm p}}(0) )^2 \rangle }.
\end{eqnarray}
At zero temperature,
\begin{equation}
\big\langle e^{i({\phi_\pm^0}(x) - {\phi_\pm^0}(0) )} \big\rangle
= \big\langle e^{\pm 2 \pi i N_\pm x/L}  \big\rangle = 1
\end{equation}
and (for $g=1$)
\begin{equation}
\big\langle {\phi_\pm^{\rm p}}(x) {\phi_\pm^{\rm p}}(0)
- ({\phi_\pm^{\rm p}}(0) )^2 \big\rangle = {\cal S}(\pm x),
\end{equation}
where ${\cal S}(x)$ is defined in Appendix B. Thus, we find
\begin{equation}
C_\pm(x) = { (\pm i / 2L) e^{\pm i \pi x/L} \over
\sin[ \pi(x \pm ia)/L] }.
\label{FL correlator}
\end{equation}
Note that (\ref{FL correlator}) has the correct periodicity
$C_\pm(x+L) = C_\pm(x)$. The same quantity
(\ref{correlator definition}) may also be calculated
directly from the chiral Fermi liquid Hamiltonian
\begin{equation}
H_\pm = \sum_k \epsilon_\pm(k) :c_\pm^\dagger(k) c_\pm(k):,
\end{equation}
normal-ordered with respect to an infinite Dirac sea as in
the Luttinger model.
Here $c_\pm^\dagger(k)$ and $c_\pm(k)$ denote creation
and annihilation operators for the right $(+)$ or
left $(-)$ branch, $\epsilon_\pm(k) = \pm vk$ are the
energies of the infinite linear branches, and
$\psi_\pm(x) = L^{-{1 \over 2}} \sum_k e^{ikx} c_\pm(k)$.
We then obtain
\begin{equation}
C_\pm(x) = {1 \over L} \sum_k e^{ikx}
\big\langle 1-c_\pm^\dagger(k) c_\pm(k) \big\rangle.
\end{equation}
The ground state momentum distribution function $n_\pm(k)$
is given by $n_+(k) =1$ for $k \leq 0$,
$n_{-}(k) =1$ for $k \geq 0$, and $n_\pm(k)=0$
elsewhere. [This may be written as $n_\pm(k)
= \theta(\mp k)$ with the understanding that the {\it full}
$k=0$ state is to be included.] Then we find
\begin{equation}
C_\pm(x) = {1 \over L} \sum_{k > 0}
e^{\pm ikx} e^{-ka},
\end{equation}
where we have included a convergence factor. This result
is identical to the form (\ref{FL correlator}) calculated
with the finite-size bosonization formula
(\ref{bosonization}).

The bosonization formula (\ref{bosonization}) may also be used to determine
the allowed eigenvalues of the charge operator $N_\pm$.
Equation (\ref{bosonization})
implies that
\begin{equation}
\psi^\dagger_\pm(x+L) = \psi^\dagger_\pm(x)
\ \! e^{\mp 2\pi i N_{\pm}/g}.
\end{equation}
Thus, periodic boundary conditions on the electron creation operators
lead to the result that the allowed
eigenvalues of $N_{\pm}$ are given by
\begin{equation}
N_\pm = n g ,
\label{fractional charge}
\end{equation}
where $n$ is any integer, which means that there exists
{\it fractionally charged}
excitations, as expected in a FQHE system.
The result (\ref{fractional charge}) also follows from the requirement
that the annihilation operators $\psi_\pm(x)$ satisfy periodic boundary
conditions.

The CLL theory (\ref{euclidian action}) is valid for any magnetic field
strength within a given fractional plateau (of the form $1/q$ with $q$ odd)
that makes the
bulk quantum Hall fluid incompressible, but does not distinguish between these
different possible magnetic fields. This is because the action
(\ref{euclidian action})
does not know whether the edge it describes is at the boundary of a Hall bar,
where the precise location within a Hall plateau is unimportant, or is at the
boundary of a quantum Hall droplet or an antidot piercing
an otherwise uniform fluid,
where there are additional mesoscopic effects associated
with the actual position
within the plateau. To account for these mesoscopic effects in the
antidot system, we couple to an AB flux $\Phi$ by adding a term
\begin{equation}
\delta {\cal L}_\pm = {1 \over c} j_\pm A
\label{flux coupling}
\end{equation}
to the CLL Lagrangian (\ref{lagrangian})
where $j_\pm(x)$ is the
one-dimensional current density associated with
$\rho_\pm(x)$, as defined through the continuity equation
$ -e \partial_t \rho_\pm + \partial_x j_\pm = 0$. Using (\ref{chiral
equations of motion})
we obtain a bosonized expression for the current density,
$j_\pm = \pm {e \over 2 \pi} \partial_t \phi_\pm$.
The vector potential in (\ref{flux coupling}) is given by $A = \Phi/L$.
The flux couples only to the zero-modes, and results in the Hamiltonian
\begin{equation}
H_\pm = {\pi v \over g L} \big(N_\pm \pm g \varphi \big)^2
+ \sum_{k} \theta(\pm k) v |k|
a_k^\dagger a_k ,
\label{hamiltonian with flux}
\end{equation}
where $\varphi \equiv \Phi / \Phi_0$ is the dimensionless flux.

\section{scaling theory}

In Section II we studied the AB effect in the integral quantum Hall regime,
where the edge
states are chiral Fermi liquids, by using the B\"uttiker-Landauer formula,
which is valid
for all values of the antidot tunneling amplitudes. In the fractional
regime, where
the edge states are CLLs, the strong electron correlation dramatically
changes the physics
of the tunneling process itself, as emphasized by Wen \cite{Wen reviews}
and also by Kane
and Fisher \cite{KF LL} and Matveev {\it et al.} \cite{Matveev etal} in the
context of the
one-dimensional interacting electron gas. Whereas electron tunneling
between FQHE edge states
is inhibited because an electron added to an edge
state is not properly correlated with the others already there, tunneling
of fractionally charged
quasiparticles, when allowed, is actually enhanced and may become divergent
at low temperatures.
Quasiparticle tunneling is difficult to treat theoretically because of this
nonperturbative aspect.

The renormalization group (RG) has played a central role in
the theory of the chiral and
nonchiral Luttinger liquids, and we now briefly review
its connection with the divergent
quasiparticle tunneling between FQHE edge states in
a quantum point contact. The divergence
reflects the fact that the quasiparticle tunneling
operator is relevant in the RG sense
\cite{KF LL,Moon etal}, and in the zero-temperature limit CLL
theory predicts that in the
quantum-point-contact geometry the Hall fluid with
interedge tunneling is an {\it insulator}.
If, by adjusting the magnetic field or a gate
voltage or both, a condition of complete destructive
interference is achieved that prevents the aforementioned
tunneling, a {\it resonance} peak in
the two-terminal conductance will occur \cite{resonance footnote}.
However, this on-resonance
fixed-point is clearly unstable and the system will try to flow to
the stable insulating fixed-point.
The situation is entirely analogous to
a second-order
phase transition, and here too we expect critical
exponents associated with the unstable
fixed-point. These exponents, which reflect the form
of the tunneling conductance near a
resonance, can be obtained from a RG analysis.
Because no other operators allowed by symmetry
are relevant at low temperatures, Kane and
Fisher \cite{KF LL} went further by proposing that
the entire RG trajectory from Hall conductor to insulator
will be {\it universal}, resulting
in a universal lineshape near and away from resonance.
The universal lineshape for the $g=1/3$
case has been calculated exactly by Fendley, Ludwig, and
Saleur \cite{Fendley etal} using
the thermodynamic Bethe ansatz.

The scaling behavior in the antidot geometry differs
in two important ways. First, although the
RG flow equations are the same here
as in the quantum-point-contact geometry, their
physical implications are different. Recall that
in the quantum-point-contact system it is
assumed, and reasonably so, that there exist conditions
of complete destructive interference
that will cause a resonance. But in the antidot
system the origin of interference is the AB effect,
which, because of the chirality, {\it never} leads
to the necessary complete destructive interference for
any finite antidot-coupling strength (see the
discussions in Section I and Appendix A). The
two-terminal conductance on resonance, $G^*$, is
determined by the antidot-coupling
strength and the
temperature and is always less than $g e^2/h$.
The second important difference between
the two geometries is the role of the additional
scale $T_0$ in the antidot case. Whereas in
the quantum-point-contact geometry the universal RG trajectory
implies a one-parameter
universal scaling function ${\tilde G}(X)$, where
$X$ depends on the temperature, the RG
trajectory in the antidot problem, which we also
predict to be universal for sufficiently
low $T_0$, leads instead to a {\it two-parameter}
universal scaling function ${\tilde G}(X,Y)$,
where $Y$ depends on the size of the antidot and is temperature independent.
The scaling function ${\tilde G}(X,Y)$ contains all the mesoscopic
effects associated with the finite-size antidot edge state.

We turn now to a detailed RG theory of the antidot
problem.  We begin by performing a perturbative
analysis in the weak-antidot-coupling regime shown
schematically in Fig.~\ref{geometry}a. In this
case we have
\begin{equation}
S = S_0 + \delta S,
\label{total action}
\end{equation}
where $S_0 \equiv S_{\rm L} + S_{\rm R} + S_{\rm A}$ is the sum
of actions of the form
(\ref{euclidian action}) for the left moving,
right moving, and antidot edge states,
respectively, and
\begin{equation}
\delta S \equiv \sum_{m=1}^\infty \int_0^\beta d \tau \ \bigg( V_{+}(\tau)
+ V_{-}(\tau) + {\rm c.c.} \bigg)
\end{equation}
is the weak coupling between them. Here
\begin{equation}
V_\pm(\tau) \equiv {v \Gamma_\pm^{\scriptscriptstyle (m)} \over 2 \pi a}
\ e^{i m \phi_\pm(x_\pm,\tau)} e^{-im \phi_A(x_\pm,\tau)}
\label{tunneling term}
\end{equation}
describes the tunneling of $m$ quasiparticles
from an incident edge state into the antidot
edge state at point $x_\pm$ with dimensionless
amplitude $\Gamma_\pm^{\scriptscriptstyle (m)}$
\cite{KF LL}.
In the weak-antidot-coupling regime
$|\Gamma_\pm^{\scriptscriptstyle (m)}| \ll 1$.
The form of the tunneling perturbation
(\ref{tunneling term}) follows from a
generalization of the bosonization formula
(\ref{bosonization}) to particles with fractional
charge and fractional statistics; the edge-state
velocity factor $v$ is included to make the
$\Gamma_\pm^{\scriptscriptstyle (m)}$ dimensionless.
The phase factor in (\ref{bosonization}) is unnecessary here and
for simplicity will be omitted.
Because the high-energy cutoff of the
theory---the effective Fermi energy---is of order $v/a$, the magnitudes of the
$\Gamma_\pm^{\scriptscriptstyle (m)}$ are essentially
tunneling rates in units of the Fermi
energy. (Recall that $a$ is a microscopic cutoff
length taken to be of the order of the
magnetic length.)  We assume
the incident edge states or leads, described by
$S_{\rm L}$ and $S_{\rm R}$, to be macroscopic,
and we also assume for simplicity that
$|\Gamma_-^{\scriptscriptstyle (m)}|=
|\Gamma_+^{\scriptscriptstyle (m)}|$. Furthermore, it
is not necessary to explicitly include
the flux in $S_{\rm A}$ so we set $\varphi = 0$.

It is most convenient to perform the RG analysis
directly in terms of the Euclidian action
(\ref{total action}), and hence in this section
expectation values of fields refer to
their functional-integral form \cite{path integral footnote},
\end{multicols}
\begin{equation}
\big\langle \phi_\pm(x_1,\tau_1) \ \!  \phi_\pm(x_2,\tau_2) \cdots \big\rangle
= {1 \over Z_\pm}  \int {\cal D}\phi_\pm \ \phi_\pm(x_1,\tau_1) \ \!
\phi_\pm(x_2,\tau_2)
\ \! \cdots \ e^{-S_\pm},
\label{path integral}
\end{equation}
where $Z_\pm \equiv \int {\cal D}\phi_\pm \ \! e^{-S_\pm}$ is the
edge-state partition function.
The allowed field configurations in (\ref{path integral}) satisfy the
boundary conditions
\begin{eqnarray}
\phi_\pm(x+L,\tau) - \phi_\pm(x,\tau) &=& 2 \pi n g \\
\phi_\pm(x,\tau + \beta) - \phi_\pm(x,\tau) &=& 0 ,
\label{boundary conditions}
\end{eqnarray}
and the measure in (\ref{path integral}) implicitly includes a sum over the
integer winding
number $n$. Note that there is only one homotopy index here, because in the
CLL a topological
charge excitation and topological current excitation are equivalent.

For our analysis we will need the Euclidian $m$-quasiparticle propagator
\begin{equation}
G_\pm^{\scriptscriptstyle (m)}(x,\tau) \equiv
\langle e^{i m \phi_\pm(x,\tau)} e^{-im \phi_\pm(0)} \rangle
\end{equation}
for a right $(+)$
or left $(-)$ moving edge state, which at zero temperature is given by
\begin{equation}
G_\pm^{\scriptscriptstyle (m)}(x,\tau) = \bigg({\pm i\pi a/L \over \sin
[\pi (x\pm iv\tau)/L]}
\bigg)^{m^2 g} \! \! \! ,
\label{quasiparticle propagator}
\end{equation}
where we have neglected the regularization in the denominator, which is
only necessary when
$x \pm i v \tau = 0$. The expression (\ref{quasiparticle propagator}) is
calculated in
Appendix B using the finite-size bosonization formalism developed in
Section III.
Note that when $x=0$, the largest contribution to an integral of
(\ref{quasiparticle propagator}) over imaginary time comes from small
$\tau$ where we have
$G_\pm^{\scriptscriptstyle (m)}(0,\tau) \sim (1/\tau)^{2\Delta},$ where
$\Delta =
m^2 g/2$ is the local scaling dimension of $e^{i m \phi_{\pm}}$.

Consider now the correlation function
\begin{eqnarray}
\big\langle  V_+(\tau) V_+^*(0) \big\rangle
&=& { v^2  \big| \Gamma_+^{\scriptscriptstyle (m)} \big|^2 \over 4 \pi^2 a^2}
\big\langle  e^{i m \phi_{+}(x_+,\tau)} e^{-im \phi_{+}(x_+,0)} \big\rangle
\big\langle  e^{-i m \phi_{A}(x_+,\tau)} e^{im \phi_{A}(x_+,0)} \big\rangle
\nonumber \\
&=& { v^2  \big| \Gamma_+^{\scriptscriptstyle (m)} \big|^2 \over 4 \pi^2 a^2}
\ \! G_{+}^{\scriptscriptstyle (m)}(0,\tau) \ \! G_{\rm
A}^{\scriptscriptstyle (m)}(0,-\tau),
\label{VV correlation function}
\end{eqnarray}
\begin{multicols}{2}
\noindent which arises in a perturbative calculation of the full partition
function
\begin{equation}
Z = \int {\cal D}\phi_L {\cal D}\phi_R {\cal D}\phi_A \ \! e^{-S}
\label{full Z}
\end{equation}
at second order. For $\langle V_{+}(\tau) V_{+}^*(0) \rangle$---and
therefore $Z$---to be
invariant under a small increase in unit-cell size $a \rightarrow a' = b a$,
we require
$\Gamma_{+}' = b^{1-2 \Delta} \Gamma_{+}$  or $d \Gamma_{+} / d\ln b =
(1-m^2 g) \Gamma_{+}.$
An analogous analysis of $\langle V_{-}(\tau) V_{-}^*(0) \rangle$ shows
that $\Gamma_{-}$
scales identically. These leading-order flow equations,
\begin{equation}
{d \Gamma_{\pm}^{\scriptscriptstyle (m)} \over d \ell} = \big( 1-m^2 g \big)
\Gamma_{\pm}^{\scriptscriptstyle(m)},
\label{flow equations}
\end{equation}
where $\ell \equiv \ln(a'/a)$, show that quasiparticle $(m=1)$ backscattering
processes are relevant whereas electron $(m=1/g)$ backscattering is
irrelevant when $g=1/3$, as
stated above. They were first derived by Kane and Fisher \cite{KF LL} using
momentum-shell RG.

One might expect the flow equations (\ref{flow equations}) to be modified
by the finite-size
of the antidot edge state. To see that this is not so, consider the
correlation function
\begin{equation}
\big\langle  V_{+}(\tau) V_{+}^*(\tau') V_{-}(\tau'') V_{-}^*(0) \big\rangle,
\end{equation}
which appears at fourth order in $\delta S$. A Wick expansion gives {\it
local} terms as in
(\ref{VV correlation function}), and, in addition, {\it nonlocal} antidot
propagators
$G_{\rm A}^{\scriptscriptstyle (m)}(x, \tau)$ with $x=L/2$, where $L$ is
now the circumference
of the antidot edge state. However, the expression
(\ref{quasiparticle propagator}) shows that nonlocal terms scale in the
{\it same} way as
the local terms. The simplicity of this conclusion is the advantage of our
RG method: It
focuses directly on the scaling of the cutoff $a$ rather than on the
scaling of $\tau$,
which is inconvenient when $x \neq 0$. Our conclusion is that the
Kane-Fisher flow equations
(\ref{flow equations}) are valid in the antidot geometry considered here.

Exactly on resonance, defined by the condition that all the
$\Gamma_{\pm}^{\scriptscriptstyle(m)}$
are zero, we have  perfect source-drain transmission
with $G^* = g e^2/h$. Note, however, that
this {\it perfect} resonance can be experimentally
realized only by physically decoupling the
antidot from the leads---it cannot occur because
of the AB effect itself. Nonetheless, the
perfect resonance is still a fixed point solution of (\ref{flow equations}),
albeit a physically
uninteresting one. However, in the weak-antidot-coupling
regime of Fig.~\ref{geometry}a, the
bare tunneling amplitudes can be very small, in which case
the leading-order RG equations
(\ref{flow equations}) yield
\begin{equation}
\Gamma^{\scriptscriptstyle(m)}_{\pm,{\rm ren}}
= \bigg({a_{\rm ren}\over a}\bigg)^{1-m^2g}
\Gamma_{\pm}^{\scriptscriptstyle(m)}.
\label{flow solution}
\end{equation}
Here $\Gamma_{\pm}^{\scriptscriptstyle(m)}$ are the bare
tunneling amplitudes and
$\Gamma^{\scriptscriptstyle(m)}_{\pm,{\rm ren}}$ are
the renormalized coupling constants at a length
scale $a_{\rm ren} > a$. In the $g=1/3$ case we see
that the renormalized $m=1$ quasiparticle
tunneling amplitude diverges as $(a_{\rm ren}/a)^{2/3}$,
whereas all others with $m>1$ vanish as
$(a_{\rm ren}/a)^{1-m^2/3}$. However, even at zero
temperature in an infinite Hall bar,
the scaling may be cut off by the finite size $L$ of the
antidot edge state, because the
effective unit-cell size $a_{\rm ren}$ cannot become
larger than the antidot \cite{RG footnote}.
Thus, the renormalized
$m=1$ quasiparticle tunneling amplitude will not
diverge here, but it can still become
very large, of
the order of $(L/a)^{2/3}$ times the bare value,
where $L/a$ is typically $10^2$.
At finite temperature the thermal length $ L_{\rm T} = v/T$,
which is the size
of the system in the imaginary-time direction,
will also cutoff the scaling behavior.
The maximum allowed $a_{\rm ren}$ is the minimum $L_{\rm min}$ of $L$
and $L_{\rm T}$,
so the final renormalized couplings in the weak-antidot-coupling regime are
\begin{equation}
\Gamma^{\scriptscriptstyle(m)}_{\pm,\rm ren}
= \big(L_{\rm min}/a \big)^{1-m^2g}
\Gamma_{\pm}^{\scriptscriptstyle(m)}.
\label{renormalized couplings}
\end{equation}
At temperatures less than the crossover temperature
$T_0$, defined in Eqn.~(\ref{T0}),
we see that
the renormalized tunneling amplitudes are determined by
$L$ whereas above $T_0$ they
are determined by $L_{\rm T}$.

As in the conventional momentum-shell procedure,
our RG transformation is based on a
coarse graining of the system followed by a
rescaling to hold the partition function
(\ref{full Z}) fixed. The flow equations (\ref{flow equations})
therefore hold quite generally
for equilibrium and nonequilibrium properties, but, for this same
reason, do not directly
describe the scaling of the specific quantity we are
after, namely, the two-terminal linear conductance as a function of flux.
[In fact, including the AB flux $\varphi$
in the antidot action $S_{\rm A}$ simply modifies
the propagator
$G_{\rm A}^{\scriptscriptstyle (m)}(x,\tau)$
by a phase and does not change the RG equations.]
To obtain scaling equations specific to the conductance
$G$ as a function of tunneling amplitudes,
temperature, and flux, one should
perform the RG transformation keeping $G$ itself invariant.
However, near a resonance
the effect of the flux on the quantity $G$
is to simply vary the tunneling
amplitudes $\Gamma_{\pm}^{\scriptscriptstyle(m)}$.
Therefore, in our application
of (\ref{flow equations}) and (\ref{renormalized couplings})
to the study of
$G$, we can simply regard the $\Gamma_{\pm}^{\scriptscriptstyle(m)}$
as being flux-dependent and gate-voltage dependent quantities, as in
the quantum-point-contact geometry.
This can be made precise by the introduction of a resonance tuning
parameter $\delta$,
which is a function of AB flux or gate voltage
or both, and which specifies the distance
from a perfect resonance $\delta = 0$ in these units.
Note, however, that we cannot set $\delta$ equal to zero, because
there are no perfect resonances here.

We are now in a position to understand the nature of the resonances
in the antidot geometry at temperatures low enough where all tunneling
amplitudes except for the $\Gamma_{\pm}^{\scriptscriptstyle(1)}$ have
scaled to negligible values. The precise temperature $T_{\rm s}$ below
which this scaling occurs depends on the values of the bare irrelevant
couplings, $\Gamma_{\pm}^{\scriptscriptstyle(m)}$ (with $m > 1$),
and, of course, how small one requires them to be. We shall also assume
that $T_0 < T_{\rm s}$, which is physically realizable in an antidot
that is not too small. When $T$ and $T_0$ are both less than $T_{\rm s}$
the RG flow will follow a {\it univeral trajectory}, as in the
quantum-point-contact problem \cite{Moon etal}. Then there is a single
correlation length $\xi$ that diverges on resonance
$\Gamma_{\pm}^{\scriptscriptstyle(1)}=0$ with a critical exponent
determined by (\ref{flow equations}),
\begin{equation}
\xi \sim \bigg( {1 \over |\Gamma_{\pm}^{\scriptscriptstyle(1)} | }
\bigg)^{1 \over 1-g},
\end{equation}
and the conductance as a function of
$|\Gamma_{\pm}^{\scriptscriptstyle(1)}| \propto \delta $,
the thermal length $L_{\rm T}$, and the antidot size $L$ will obey
the scaling law
\begin{eqnarray}
G\big( \delta, L_{\rm T}, L \big) &=& {\tilde G}'\bigg(c_1 {L_{\rm T}
\over \xi} , c_2 {L \over \xi} \bigg) \nonumber \\
&=& {\tilde G}\bigg(c_1 {\delta \over T^{1-g}}, c_2 {\delta \over T_0^{1-g}}
\bigg),
\label{scaling law}
\end{eqnarray}
where $c_1$ and $c_2$ are nonuniversal constants and
${\tilde G}(X,Y)$ is a
two-parameter
universal scaling function. In the second form in
(\ref{scaling law}) we have expressed
the antidot size in terms of the crossover temperature $T_0$ using (\ref{T0}).
Near the resonance, (\ref{scaling law}) expresses the conventional
assumption about
finite-size scaling near a critical point, in both the
real-space and imaginary time directions.
Much less trivial is our assumption, following a similar one
by Kane and Fisher \cite{KF LL},
that (\ref{scaling law}) holds for {\it all} values of $\delta$
over which
$|\Gamma_{\pm}^{\scriptscriptstyle(1)}| \propto \delta $.
The universal scaling function ${\tilde G}(X,Y)$ will be valid as long
as the temperature is low enough so that the corrections to
scaling from the irrelevant operators are small and as long as the
resonances are narrow enough so that the linear relation
$|\Gamma_{\pm}^{\scriptscriptstyle(1)}| \propto \delta$
applies.

In the limit $L \rightarrow a$ the antidot system
becomes equivalent to quantum point contact
with a single miscoscopic impurity providing the
momentum transfer to the lattice necessary
for tunneling. The crossover from CLL
power-law behavior to nearly Fermi-liquid-like scaling
caused by mesoscopic effects
that we discuss in detail in the next section
does not occur in this limit because $T_0$ is pushed up to
the high-energy cutoff $ T_{\rm F} \equiv v/a$, the effective Fermi
temperature, beyond which CLL is invalid and
the FQHE does not occur. Therefore in the $Y \rightarrow 0$
limit our two-parameter
scaling function ${\tilde G}(X,Y)$ reduces to the
one-parameter function
${\tilde G}(X)$
defined by Kane and Fisher\cite{KF LL} and calculated
by Moon {\it et al.} \cite{Moon etal}
and by Fendley, Ludwig, and Saleur \cite{Fendley etal}:
\begin{equation}
{\tilde G}(X,0) = {\tilde G}(X).
\end{equation}
An explicit form for ${\tilde G}(X,Y)$, valid for $X \gg 1$, may be
obtained from perturbation theory in the strong-antidot-coupling regime,
but will not be needed here.

The scaling law (\ref{scaling law}) shows that there is a temperature scale
$T_1$ determined by the bare quasiparticle tunneling amplitude
$\Gamma_{\pm}^{\scriptscriptstyle(1)}$ below which the system
is always in the strong-antidot-coupling regime. This means that
the resonances are never perfect, and at temperatures less than $T_1$
the on-resonance conductance $G^* \ll g e^2/h$. At temperatures
above $T_1$ a weak-antidot-coupling regime is of course possible for
$|\Gamma_{\pm}^{\scriptscriptstyle(m)}| \ll 1$, and in this case
\begin{equation}
G^* = \big[ 1 - {\cal O}(\Gamma^4) \big] g {e^2 \over h}
\approx g {e^2 \over h}.
\end{equation}
If the antidot system starts in the strongly coupled regime, by an appropriate
choice of gate voltages, it will stay in this regime throughout the
experimentally relevant
ranges of temperature and magnetic field, because the $m=1$ quasiparticle
backscattering
process (which would be relevant in the RG sense) is not allowed in this
edge-state
configuration and only electrons can tunnel\cite{Wen pert theory}. The
strong-antidot-coupling
regime therefore admits a perturbative treatment \cite{Wen pert theory}, to
which we now turn.

\section{Aharonov-Bohm effect in the strong-antidot-coupling regime}

The current $I$ passing between edge states $L'$ and $R'$ as a function
of their potential difference $V$ may be calculated for small 
energy-independent tunneling
amplitudes $\Gamma_i \ (i=1,2)$, which for simplicity are taken to
be equal apart from dynamical and AB phase factors,
\begin{eqnarray}
\Gamma_1 &=& \Gamma e^{i \pi ({\mu \over \Delta \epsilon} + \varphi)}
\nonumber \\
\Gamma_2 &=& \Gamma e^{-i \pi ({\mu \over \Delta \epsilon} + \varphi)}.
\end{eqnarray}
Here $\varphi \equiv \Phi/\Phi_0$, where $\Phi_0\equiv hc/e$ is the
flux quantum, and $\mu \equiv (\mu_{\rm L} + \mu_{\rm R})/2$ is
the mean electrochemical potential. These phases account for the total
phase
$\theta \equiv \oint_\epsilon d{\bf l} \cdot
({\bf p} + {e \over c} {\bf A})$
accumulated by an electron with energy $\mu$ after one complete
clockwise orbit around the antidot.
The Hamiltonian is $H = H_0 + \delta H$, where $H_0 = H_{\rm L}
+ H_{\rm R}$ is a sum of Hamiltonians of the form (\ref{hamiltonian}) and
\begin{equation}
\delta H = \Gamma_1 B_1 + \Gamma_2 B_2 + \Gamma_1^* B_1^\dagger
+ \Gamma_2^* B_2^\dagger,
\end{equation}
where $B_i \equiv \psi_L(x_i) \psi_R^\dagger(x_i)$
is an electron tunneling operator acting at point $x_i$.
Because the edge states $L'$ and $R'$ are assumed to be
infinite, the additional c-number phase factor in
(\ref{bosonization}) is not needed here.
The current $I(t) \equiv -e \langle {\dot N}_{\rm L} \rangle$
to first order in $\delta H$ is given by
\begin{equation}
I(t) = i e \int dt' \ \theta(t-t') \ {\rm Tr} \ \rho_0
[ \partial_t {\tilde N}_{\rm L}(t) , {\tilde{\delta H}}(t') ],
\label{linear response}
\end{equation}
where
$\rho_0 \equiv e^{- \beta K_0}/{\rm Tr} \ \! e^{- \beta K_0}$,
\begin{equation}
K_0 \equiv H_0 - \mu_{\rm L} N_{\rm L} - \mu_{\rm R} N_{\rm R},
\end{equation}
and where (\ref{linear response}) is written in the usual interaction
representation ${\tilde O}(t) \equiv e^{i H_0 t} O e^{-i H_0 t}$.
It is convenient, however, to work in the $K_0$-representation
defined by
$O(t) \equiv e^{i K_0 t} O e^{-i K_0 t}$.
We find ${\tilde B}_i(t) = B_i(t) e^{iVt}$,
where $V \equiv (\mu_{\rm R} - \mu_{\rm L})/e$ is the applied voltage.
Then (\ref{linear response}) leads to
\end{multicols}
\begin{equation}
I = -2 |\Gamma|^2 \ \! {\rm Im} \ \! \bigg[
X_{11}(\omega) + X_{22}(\omega)
+e^{2 \pi i ({\mu \over \Delta \epsilon} + \varphi)} \ \! X_{12}(\omega)
+e^{-2\pi i ({\mu \over \Delta \epsilon} + \varphi)} \ \! X_{21}(\omega)
\bigg]_{\omega = eV},
\label{IV relation}
\end{equation}
where $X_{ij}(\omega)$ is the Fourier transform of
\begin{equation}
X_{ij}(t) \equiv -i \theta(t) \langle [B_i(t), B_j^\dagger(0)] \rangle.
\end{equation}
This response function can be calculated using bosonization techniques
and the result for filling factor $g = 1/q$ is
\begin{equation}
X_{ij}(t) = - \theta(t) { a^{2q-2} \over 2 \pi^2} \ {\rm Im} \ \!
{ (\pi / L_{\rm T} )^{2q}
\over \sinh^q [\pi (x_i-x_j+vt+ia)/L_{\rm T}]
\sinh^q [\pi (x_i-x_j-vt-ia)/L_{\rm T}] } ,
\label{response function}
\end{equation}
\begin{multicols}{2}
\noindent where
$L_{\rm T} \equiv v/T$ is the thermal length.
When $q=1$, (\ref{response function}) is the response function
for noninteracting chiral electrons.
From (\ref{response function}) we see that $X_{11}=X_{22}$ and
$X_{12}=X_{21}$, so (\ref{IV relation}) may be written as
\begin{equation}
I = -4 |\Gamma|^2 \ \! {\rm Im} \ \! \bigg[
X_{11}(\omega)  + \cos[2 \pi ({\textstyle {\mu \over \Delta \epsilon}}
+ \varphi )]
 \ \! X_{12}(\omega)
\bigg]_{\omega = eV}.
\label{simplified IV relation}
\end{equation}
Thus it is sufficient to calculate the imaginary part of
$X_{ij}(\omega)$, which we shall do below.

Each term $X_{ij}$ in (\ref{IV relation}) corresponds to a process
occurring
with a probability proportional to $|\Gamma_i \Gamma_j|$.
The {\it local} terms $X_{11}$ and $X_{22}$ therefore
describe {\it independent} tunneling at $x_1$ and $x_2$,
respectively, whereas the
{\it nonlocal} terms $X_{12}$ and $X_{21}$ describe {\it coherent}
tunneling through both antidot constrictions.
The AB phase naturally couples only to the latter.
We shall see that the local contributions behave exactly like the
tunneling current in a quantum point contact.
The AB effect, however, is a consequence of the nonlocal terms, and
we shall show that there are new
non-Fermi-liquid phenomena associated with these terms
that are directly accessible to experiment.

The required Fourier transform may be calculated by contour integration.
Here we shall present the calculation for the case $g=1/3$ (the other
cases follow similarly).  Because the factor multiplying $\theta(t)$
in (\ref{response function}) is odd under $t \rightarrow -t$, the
imaginary part of $X_{ij}(\omega)$ may be written as

\end{multicols}

\begin{equation}
{\rm Im} \ X_{ij}(\omega) = - {a^4 \pi^3 T^5 \over 8 v^6}
\ {\rm Re} \int_{-\infty}^\infty ds
\ {e^{i \omega s / \pi T} - e^{-i \omega s / \pi T} \over
\sinh^3(s + \pi d/L_{\rm T} + i\eta)
\sinh^3(s - \pi d/L_{\rm T} + i\eta)},
\label{contour integral}
\end{equation}
where $d \equiv |x_i - x_j|$ and $\eta$ is a positive infinitesimal.
The local response functions $X_{11}$
and $X_{22}$ correspond to $d=0$, whereas the nonlocal ones $X_{12}$
and $X_{21}$ correspond to $d=L/2$, where $L$ is the circumference
of the antidot edge state. When $d \neq 0$ there are 3rd order poles at
\begin{equation}
s = \pm {\pi d \over L_{\rm T}} + i n \pi - i \eta,
\end{equation}
where $n$ is an integer.
The integral of the first and second term in (\ref{contour integral})
can be calculated by closing
the integration contour in the upper-half-plane and lower-half-plane,
respectively. One can show that the contributions from all the poles
except $n=0$ cancel, leaving
\begin{eqnarray}
{\rm Im} \ X_{ij}(\omega) = {a^4 \pi^2 \over 8 v^6}
{T^3 \over \sinh^3(2 \pi d / L_{\rm T}) }
\bigg\lbrace
&\bigg[ & V^2+ 4\pi^2 T^2 \bigg(1-
3 \coth^2(2 \pi d / L_{\rm T}) \bigg) \bigg]
\sin\bigg({Vd \over v}\bigg) \nonumber \\
&+& 6 \pi V T \coth(2 \pi d / L_{\rm T})
\cos\bigg({Vd \over v}\bigg) \bigg\rbrace .
\label{transform}
\end{eqnarray}
The case where $d=0$ may then be obtained by taking the
$d \rightarrow 0$
limit of this expression. In the $g=1/3$ case one has to expand
up to third order in $d$ (all lower orders cancel exactly).

The response function (\ref{transform}) evidently displays a crossover
behavior when the thermal length $L_T$ becomes less than $|x_i-x_j|$.
The finite size of the antidot therefore provides
an important temperature scale (\ref{T0}).
For example, a Fermi velocity $v$ of
$10^6 \ {\rm cm/s}$ and circumference $L$ of
$1 \ \mu {\rm m}$ yields $T_0 \approx 25 {\rm mK}.$
Note that
$T_0$ is closely related to the energy level spacing
$\Delta \epsilon$
for noninteracting electrons with linear dispersion
in a ring of circumference $L$,
\begin{equation}
T_0= {\Delta \epsilon \over 2\pi^2}.
\end{equation}

The current can generally be written as
\begin{equation}
I = I_0 + I_{\rm AB} \cos\bigg[ 2 \pi \bigg(
{\mu \over \Delta \epsilon} + \varphi \bigg) \bigg] ,
\label{I separation}
\end{equation}
where $I_0$ is the \lq\lq background\rq\rq current resulting from the local
terms and $I_{\rm AB}$ is the AB current resulting from the
nonlocal terms.
When the voltage is applied symmetrically about an antidot
energy level, $\cos[2\pi({\mu \over \Delta \epsilon} + \varphi)]
= \cos(2 \pi \varphi)$.

The exact current-voltage relation for the $g=1/3$ chiral Luttinger
liquid is
\begin{equation}
I_0 = {|\Gamma|^2 a^4  \over 120 \pi v^6}
\bigg( 64 {\pi}^4 VT^4+
20 {\pi}^2 T^2 V^3 + V^5 \bigg),
\label{I0}
\end{equation}
and
\begin{equation}
I_{\rm AB}= - {|\Gamma|^2 a^4 \pi^2 \over v^6}
{ T^3 \over{\sinh^3(T/T_0)}} \bigg\lbrace
\bigg[ V^2+ 4\pi^2 T^2 \bigg(1-3 \coth^2(T/T_0) \bigg) \bigg]
\sin\bigg({VL \over 2v}\bigg) + 6 \pi V T \coth(T /T_0)
\cos\bigg({VL \over 2v}\bigg) \bigg\rbrace .
\label{IAB}
\end{equation}
In the limit $L \rightarrow 0$, $I_{\rm AB}$ always reduces to $I_0$.
The $\sin(VL/2v)$ and $\cos(VL/2v)$ factors have a period in $V$
equal to twice the level spacing $\Delta \epsilon$,
as expected (see Section II).
The AB conductance for $q=3$ is
\begin{equation}
G_{\rm AB} = -{2 \pi^3 | \Gamma |^2 a^4 \over v^6}
{ T^4 \over \sinh^3(T/T_0)} \bigg\lbrace 3 \coth \bigg({T \over T_0}\bigg)
+ \bigg({T \over T_0}\bigg) \bigg[ 1- 3 \coth^2 \bigg({T \over T_0}\bigg)
\bigg] \bigg\rbrace,
\label{GAB}
\end{equation}
\begin{multicols}{2}
\noindent which is shown in Fig.~\ref{G} along with the corresponding
chiral Fermi liquid result (\ref{FL GAB}).

We now summarize our results for general $q$. We shall for convenience
summarize the transport properties as a function of temperature for
fixed voltage, first for $V \ll T_0$ and then for $V \gg T_0$.

\subsection{Low voltage regime: $V \ll T_0$}

There are three temperature regimes here. When
$T \ll V \ll T_0$, both $I_0$ and $I_{\rm AB}$
are temperature independent but have nonlinear
behavior, varying with voltage as
\begin{equation}
I \propto V^{2q-1}.
\label{nonlinear IV}
\end{equation}

When the temperature
exceeds $V$, the response becomes linear.
When $V \ll T \ll T_0$, both $G_0$ and $G_{\rm AB}$ vary with
temperature as
\begin{equation}
G \propto \bigg({T \over T_F}\bigg)^{2q-2},
\label{low-temperature G}
\end{equation}
where $T_{\rm F} \equiv v/a$ is an effective Fermi temperature.
This temperature dependence shows that in the strong-antidot-coupling
regime the renormalization of the electron tunneling amplitudes is
not cut off by the finite size $L$ of the antidot edge state, but
only by the thermal length $L_{\rm T}$.

At a temperature near $T_0$, we find that $G_{\rm AB}$ for the CLL
displays a pronounced maximum, also in striking contrast to a Fermi
liquid [see Fig.~\ref{G}].

Increasing the temperature further we cross over into the
$V \ll T_0 \ll T$ regime where $G_0$ scales as in
(\ref{low-temperature G}), but
\begin{equation}
G_{\rm AB} \propto \bigg({T \over T_0} \bigg)
\bigg({T \over T_{\rm F}} \bigg)^{2q-2} e^{-qT/T_0}.
\label{high-temperature G}
\end{equation}
Thus $G_{\rm AB}$ exhibits a crossover from the well-known
$T^{2q-2}$ Luttinger liquid behavior to a new scaling behavior
which is much closer to a chiral Fermi liquid $(q=1)$.
Careful measurements in this experimentally accessible regime should
be able to distinguish between a Fermi liquid  and this
predicted nearly Fermi-liquid temperature dependence.

\subsection{High voltage regime: $V \gg T_0$}

Again there are three temperature regimes. For the lowest
temperatures $T \ll T_0 \ll V$, the response is again
temperature independent and
nonlinear. The direct term varies with voltage according to
\begin{equation}
I_0 \propto V^{2q-1},
\end{equation}
as in the lowest temperature,
low voltage regime. However, the flux-dependent part of the current
is now much more interesting, involving power-laws times Bessel
functions of the ratio $V/2 \pi T_0 = \pi V / \Delta \epsilon$.
As an example, for the case $q=3$
we find in this regime
\begin{eqnarray}
I_{\rm AB} = {4 e |\Gamma|^2 a^4 \over \pi v L^5}
\bigg\lbrace
&\bigg[& 3 - \bigg({V\over 2\pi T_0}\bigg)^2 \bigg]
\sin \bigg({V \over 2\pi T_0} \bigg) \nonumber \\
&-& \bigg[{3V\over 2\pi T_0}\bigg] \cos \bigg({V \over 2\pi T_0}\bigg)
\bigg\rbrace,
\end{eqnarray}
which is shown in Fig.~\ref{I} along with the chiral Fermi
liquid result (\ref{FL IAB}) at zero temperature.
Note that in this low-temperature regime
\begin{equation}
I_{\rm AB} \propto \bigg( {V \over 2 \pi T_0} \bigg)^{5/2}
 J_{5/2}\bigg( {V \over 2 \pi T_0} \bigg),
\label{bessel}
\end{equation}
where $J$ is a Bessel function of the first kind,
a result also obtained by Chamon {\it et al.} \cite{Chamon etal}
for the douple point-contact geometry.

As the temperature is increased further to
$T_0 \ll T \ll V$, we find a crossover
to an interesting high-temperature nonlinear regime.
Here $I_0 \propto V^{2q-1}$ as before, but now
\begin{equation}
I_{\rm AB} \propto \bigg({T \over T_0}\bigg)^q e^{-qT/T_0} V^{q-1}
\sin \bigg({V \over 2 \pi T_0} \bigg).
\label{high-temperature nonlinear IAB}
\end{equation}
Therefore, the nonlinear response at fixed temperature
can also be used to distinguish between Fermi liquid and
Luttinger liquid behavior, even at relatively high temperatures.

When the temperature exceeds V, the response finally
becomes linear.
When $T_0 \ll V \ll T$, $G_0$ scales as in
(\ref{low-temperature G}) whereas $G_{\rm AB}$ scales
as in (\ref{high-temperature G}).
Thus at  high temperatures the low- and high-voltage
regimes behave similarily.

\section{persistent current in a chiral Luttinger liquid}

In the previous section we have been discussing the transport properties
of an edge state that occurs at the boundary of a quantum Hall fluid
pierced by an antidot potential. In this section we shall discuss a
non-Fermi-liquid mesoscopic
property of the edge current occurring
at this same type of antidot boundary or at the boundary of a
FQHE droplet confined in a quantum dot.

In a macroscopic edge state, an equilibrium edge current exists even
in the absence of an AB flux or twisted boundary conditions.
The magnitude of this current is universal and in the
absence of disorder is given by \cite{Geller and Vignale}
\begin{equation}
I_{\rm edge} = g {e\omega_{\rm c} \over 4 \pi}
+ {e {\tilde \epsilon}_{\rm qh} \over 2 \pi},
\label{universal edge current}
\end{equation}
where $\omega_{\rm c}$ is the cyclotron
frequency and
${\tilde \epsilon}_{\rm qh}$ is the {\it proper} quasihole energy
(defined at fixed density)
of the Laughlin state at filling factor $g = 1/q$.

We now couple the edge state to an AB flux
$\varphi \equiv \Phi /\Phi_0$.
The grand-canonical partition function of the mesoscopic edge state
factorizes into a zero-mode contribution,
\begin{equation}
Z^0 = \sum_{n=-\infty}^\infty e^{- g \pi^2 (T_0/T)
(n-\varphi)^2},
\label{zero-mode partition function}
\end{equation}
which depends on $\Phi$, and an irrelevant
flux-independent contribution
from the nonzero-modes.
Here $T_0$ is again given by Eqn. (\ref{T0}).
Note that if $N_\pm$ were restricted to be an integer
then the period of these equilibrium AB oscillations would be
$\Phi_0/g$.
The allowed fractionally charged excitations
(\ref{fractional charge}) are therefore responsible for
restoring the AB period to $\Phi_0$, as is well-known in
other contexts \cite{AB period}.

The edge current induced from the additional
flux $\Phi$ is
\begin{equation}
I \equiv - {\partial \Omega \over \partial \Phi}
= {2 \pi  T \over \Phi_0 } \sum_{n=1}^\infty
(-1)^n { \sin(2 \pi n \varphi) \over
\sinh(n q T / T_0)},
\label{persistent current}
\end{equation}
where $\Omega$ is the grand-canonical potential.
At zero temperature, this {\it chiral persistent current}
has an amplitude (with units now restored)
\begin{equation}
{\bar I} = g {e v \over L},
\label{persistent current magnitude}
\end{equation}
where $L$ is the length of the edge state.
Note that
${\bar I}$ is renormalized by the
electron-electron interactions in precisely the
same way as in a nonchiral Luttinger liquid \cite{Loss}.
At temperatures $T \gg T_0$ the amplitude decays as
\begin{equation}
{\bar I} \approx g {e v \over L} e^{-q T/T_0}.
\end{equation}
Because these persistent currents are chiral, there is no
backscattering from impurities and hence no amplitude
reduction from weak disorder.
The temperature dependence of the orbital magnetic response
of a FQHE edge state may therefore be another ideal system to observe
non-Fermi-liquid mesoscopic behavior.

\section{discussion}

We have studied the tunneling through an edge state formed around an
antidot in the fractional quantum Hall effect regime using chiral
Luttinger liquid theory. Our analysis has shown that the 
quantum-point-contact and antidot geometries are considerably different:
(i) First, mesoscopic effects are important in the antidot geometry 
when the thermal length becomes smaller than the size of the antidot,
and this leads to a crossover from the power-law tunneling characteristics
normally associated with a Luttinger liquid to a Fermi-liquid-like
scaling. Therefore, mesoscopic effects in a Luttinger liquid can mimic   
Fermi-liquid behavior. This has been demonstrated explicitly for the
strong-antidot-coupling case, but it is clear that a similar crossover
must occur for all values of the antidot tunneling amplitudes. 
(ii) The second difference is that because of the unusual nature of
the Aharonov-Bohm interference process in a chiral system, there
are never perfect resonances in the antidot system, even at zero
temperature. This means that at low enough temperatures the system
will always be in the strong-antidot-coupling regime, for all values
of the Aharonov-Bohm flux and gate voltages. The sharp non-Fermi-liquid
resonance studied in the quantum-point-contact geometry, having
a width varying with temperature as $T^{1-g}$, is therefore not expected
in the antidot geometry at the lowest temperatures.

We have also identified a new experimentally realizable regime, 
the strong-antidot-coupling regime, 
where striking non-Fermi-liquid mesoscopic transport phenomena 
are predicted. This regime is ideal for experimental investigation 
because the exact current-voltage relation is known 
[for example, Eqns.~(\ref{I separation}) through (\ref{IAB})],
and the low-temperature
crossover from weak-antidot-coupling to strong-antidot-coupling
does not complicate the analysis.  
If, by an appropriate choice of gate voltages, 
the antidot system starts in the strongly coupled regime, 
then it will stay in this regime throughout the experimentally 
relevant ranges of temperature and magnetic field.

Finally, we have predicted new mesoscopic edge-current oscillations 
or \lq\lq chiral persistent currents\rq\rq
that have a universal non-Fermi-liquid
temperature dependence and may be another means to observe a
chiral Luttinger liquid.

\acknowledgements

This work has been supported by NSERC of Canada.
It is a pleasure to thank
Claudio Chamon,
Chris Ford,
John Franklin,
Denise Freed,
Steve Girvin,
Vladimir Goldman,
Duncan Haldane,
Jung Hoon Han,
George Kirczenow,
Ilari Maasilta,
Kyungsun Moon,
David Thouless,
and Ulrich Z\"ulicke
for useful discussions.
M.G. would like to acknowledge the kind hospitality of the Aspen Center
for Physics and the University of Washington where some of this work was
carried out.

\appendix

\section{quantum mechanics of the noninteracting chiral electron gas}

In this appendix we summarize properties of the chiral Fermi liquid that are
needed in the body of the paper. In the presence of a dimensionless AB flux
$\varphi \equiv \Phi/\Phi_0$, the single-particle Hamiltonian for right $(+)$
or left $(-)$ movers is
\begin{equation}
H = \pm v( p - {\textstyle{2 \pi \varphi \over L}}),
\end{equation}
with eigenfunctions $ \phi_\pm^n(x) = L^{-{1 \over 2}} e^{2 \pi i n x /L}$ and
eigenvalues $\pm (n-\varphi)\Delta \epsilon$, where $n$ is an integer,
$\Delta \epsilon \equiv 2 \pi v/L$ is the level spacing,
and where periodic boundary conditions on a line of length $L$ have been used.

The retarded Green's function
\begin{equation}
G_\pm^{\rm R}(x,t)\equiv -i\langle \{\psi_\pm(x,t),\psi_\pm^\dagger(0)\}
\rangle \theta (t)
\end{equation}
for a free chiral electron gas has an especially simple form, namely
\begin{equation}
G_\pm^{\rm R}(x,t) = -i \ \!  \theta (t) \ \! e^{\pm 2 \pi i \varphi vt/L}
\sum_{n=-\infty}^\infty \delta(x \mp vt + nL).
\label{FL retarded G(x,t)}
\end{equation}
An electron added to the system therefore propagates ballistically with
a velocity $v$ and with no dispersion.
The Fourier transform of $G_\pm^{\rm R}(x,t)$ is
\begin{eqnarray}
G_\pm^{\rm R}(x, \omega)
&=& {1 \over L} \sum_{n=-\infty}^\infty {e^{2 \pi i n x / L} \over \omega \mp
(n-\varphi) \Delta \epsilon + i \eta} \label{first form}  \\
&=& - {i \over v}
\sum_{n=-\infty}^\infty \theta\big(\pm (x+nL) \big)
e^{2 \pi i ( {\omega \over \Delta \epsilon} \pm \varphi)({x \over L} + n)} .
\label{second form}
\end{eqnarray}
The second form (\ref{second form}) follows from (\ref{first form}) by an
application
of the Poisson summation formula, or from (\ref{FL retarded G(x,t)}) directly.
The second expression has a useful interpretation:
$G_\pm^{\rm R}(x, \omega)$ is proportional to the amplitude for an electron
to propagate
a distance $x$ around the ring via a direct path, during which it acquires
a phase
\begin{equation}
\exp\bigg[ i\bigg( {2\pi \omega \over \Delta \epsilon} \pm 2 \pi \varphi
\bigg){x\over L} \bigg],
\label{accumulated phase}
\end{equation}
plus the amplitude to propagate via any number of windings around the ring
with a given chirality.
The first term in (\ref{accumulated phase}) is the dynamical phase, whereas the
second term is the chirality-dependent AB phase.

The total amplitude to propagate at frequency $\omega$ from point 1 to
point 2 on the
ring shown in Fig.~\ref{ring}, allowing only clockwise $(+)$ or
counterclockwise
$(-)$ motion, is proportional to
\begin{equation}
G_\pm^{\rm R}({\textstyle{L \over 2}},\omega) = {1 \over 2 v
\sin \pi({\textstyle{\omega \over \Delta \epsilon}} \pm \varphi) }.
\end{equation}
Note that the chirality enters only through the AB phase.
The transmission probability is proportional to
\begin{equation}
|G_\pm^{\rm R}({\textstyle{L \over 2}},\omega)|^2 = { 1/2v^2 \over
1 - \cos 2 \pi({\textstyle{\omega \over \Delta \epsilon}} \pm \varphi) },
\label{chiral transmission probability}
\end{equation}
which possesses transmission resonances when
$ {\textstyle{\omega \over \Delta \epsilon}} \pm \varphi$ is integral,
but never exhibits completely destructive interference.

For completeness we also give expressions for the spectral function
$A_\pm(k,\omega) \equiv - 2 \ \! {\rm Im} \ \! G_\pm^{\rm R}(k,\omega)$
and density of states
$N_\pm(\omega) \equiv \int_{-\infty}^\infty {dk \over 2 \pi} A_\pm(k,\omega)$
for free chiral fermions in the infinite system-size limit:
$A_\pm(k,\omega) = 2 \pi \delta (\omega \mp vk)$
and $N_\pm(\omega) = 1/v$.

\section{euclidian quasiparticle propagator}

Here we use the results of Section III to calculate the $m$-quasiparticle
propagator for a finite-size CLL with $\varphi = 0$,
as defined in Section IV. This propagator may be
written in terms of the quantized chiral scalar
field (\ref{decomposition}) as
\begin{equation}
G_\pm^{\scriptscriptstyle (m)}(x,\tau) \equiv \langle T e^{im
\phi_\pm(x,\tau)}
e^{-im \phi_\pm(0)} \rangle,
\label{operator definition of G}
\end{equation}
where $T$ is the time-ordering operator for
particles with fractional statistics $e^{i\pi m^2 g}$.
The imaginary-time equation of motion for the
phase variable $\chi_\pm$ shows that
$\phi_\pm^0(x,\tau) = \pm {2 \pi \over L}
N_\pm (x \pm i v \tau) - g \ \! \chi_\pm$.
Separating out the zero-modes we find in the zero-temperature limit that
\end{multicols}
\begin{equation}
G_\pm^{\scriptscriptstyle (m)}(x,\tau) =
\big\langle e^{im ( \phi_\pm^0(x,\tau) - \phi_\pm^0(0) ) } \big\rangle
\bigg( e^{{m^2 \over 2}[\phi_\pm^0(x,\tau),\phi_\pm^0(0)]}
e^{m^2 g {\cal S}(\pm x + iv \tau)} \theta(\tau)
+ e^{i \pi m^2 g}  e^{-{m^2 \over 2}[\phi_\pm^0(x,\tau),\phi_\pm^0(0)]}
e^{m^2 g {\cal S}(\mp x - iv \tau)} \theta(-\tau) \bigg),
\label{expanded G}
\end{equation}
where $e^{i \pi m^2 g}$ is the statistical phase of
the $m$-quasiparticle composite and
\begin{equation}
{\cal S}(x) \equiv {2 \pi \over L} \sum_{k>0} {e^{ikx}-1 \over k} e^{-ka}.
\label{sum definition}
\end{equation}
The quantity ${\cal S}(x)$ may be found by differentiating with respect to
$x$, performing the
summation, and then integrating, which yields
\begin{equation}
{\cal S}(x) = \ln \bigg( {i \pi a /L \over \sin [\pi(x+ia)/L] }\bigg)
- {i \pi x \over L}.
\label{sum}
\end{equation}
The second term in (\ref{sum}) leads to a cancellation of
the zero-mode commutators in
(\ref{expanded G}). The remaining zero-mode expectation
value, which has the form
\begin{equation}
\big\langle e^{\pm 2\pi i m N_\pm (x\pm i v\tau)/L} \big\rangle
= { \sum_n e^{-\beta \pi v g n^2/L} e^{\pm 2\pi imng (x\pm i v\tau)/L}
\over  \sum_n e^{-\beta \pi v g n^2/L} },
\end{equation}
is equal to unity in the zero-temperature limit, so the final result is
\begin{equation}
G_\pm^{\scriptscriptstyle (m)}(x,\tau)
=\bigg({\pm i\pi a/L \over \sin[\pi (x\pm iv\tau\pm ia \ \! {\rm sgn} \ \!
\tau)/L]}
\bigg)^{m^2 g} \! \! \! \! ,
\label{final form}
\end{equation}
where the branch cut has been chosen to cancel the statistical phase.

\begin{multicols}{2}

\end{multicols}

\begin{figure}
\caption{Mesoscopic ring. $\theta_\pm$ is the phase shift subjected to an
electron of energy $\epsilon$ after a complete orbit in the clockwise $(+)$
or counterclockwise $(-)$ direction.}
\label{ring}
\end{figure}

\begin{figure}
\caption{Aharonov-Bohm effect geometry in the (a) weak-antidot-coupling and (b)
strong-antidot-coupling regimes. In both cases the arrows denote the direction
of currents and the dashed lines represent weak tunneling processes.}
\label{geometry}
\end{figure}

\begin{figure}
\caption{Zero-temperature Aharonov-Bohm resonances in the chiral
Fermi liquid with weak coupling ($\Gamma = 0.2$, solid curve),
intermediate coupling ($\Gamma = 0.5$, dashed curve), and with
strong coupling ($\Gamma = 0.9$, dotted curve). The source-drain
conductance $G$ is given in units of $e^2/h$ and the flux is in
units of the flux quantum $\Phi_0$.}
\label{AB resonances}
\end{figure}

\begin{figure}
\caption{Temperature dependence of $G_{\rm AB}$ for the cases $g=1$ (dashed
curve) and $g=1/3$ (solid curve). Both curves are normalized to have unit
amplitude at their respective maxima.}
\label{G}
\end{figure}

\begin{figure}
\caption{Nonlinear IV curve for the the cases $g=1$ (dashed curve) and $g=1/3$
(solid curve). The current is in arbitrary units and
$V_0 \equiv \Delta \epsilon/ \pi$.}
\label{I}
\end{figure}


\begin{references}
\bibitem{tunneling amplitude footnote} The antidot tunneling amplitudes
$\Gamma_{\pm}=i \Gamma$ are chosen to be purely imaginary so that in the
$\Gamma \rightarrow 0$ limit the electron suffers no phase shift at the
tunneling points.
\bibitem{path integral footnote} We shall not attempt a derivation of
(\ref{path integral}) here, which is complicated by the fact that the nonlocal
commutation relation (\ref{full commutation relation}) invalidates the usual
construction of a basis of field eigenstates as products of simultaneous
eigenstates at each point.
\bibitem{RG footnote} In the strong-antidot-coupling regime, however,
the renormalization of the tunneling amplitudes is {\it not} cut off
by $L$, but only by the finite temperature (or finite voltage).
This is because in this regime the geometry is effectively that of
two infinite edge states as shown in Fig.~\ref{geometry}b.
\end{references}
\end{document}